\documentclass[pre,onecolumn,amssymb,showkeys,superscriptaddress,notitlepage]{revtex4-1}
\pdfoutput=1
\usepackage{graphicx}
\usepackage{color}
\usepackage{epsfig}
\usepackage{latexsym}
\usepackage{bm}
\usepackage{ulem}
\usepackage{color}
\usepackage{enumerate}
\usepackage{enumitem}
\usepackage{dcolumn}
\usepackage{amsmath,amssymb}    % need for subequations
\usepackage{bm} 
\usepackage{hyperref}
\usepackage{latexsym}
\usepackage{color}
\usepackage{subfigure}
\usepackage{fancyhdr}
\def\beq{\begin{equation}}
\def\eeq{\end{equation}}

\def\Q{\mbox{\sffamily\bfseries Q}}

\def\lsim{\:\raisebox{-0.5ex}{$\stackrel{\textstyle<}{\sim}$}\:}
\def\gsim{\:\raisebox{-0.5ex}{$\stackrel{\textstyle>}{\sim}$}\:}
\def\beq{\begin{equation}}                           
\def\eeq{\end{equation}}                           
\def\bea{\begin{eqnarray}}                           
\def\eea{\end{eqnarray}}        

\begin{document}

% Use the \preprint command to place your local institutional report
% number in the upper righthand corner of the title page in preprint mode.
% Multiple \preprint commands are allowed.
% Use the 'preprintnumbers' class option to override journal defaults
% to display numbers if necessary
%\preprint{}

%%%%%%%%%%%%%%%%%%%%%%%%%%%%%%%%%%%%%%%%%%%%%%%%%%%
%				TITLE & ABSTRACT
%%%%%%%%%%%%%%%%%%%%%%%%%%%%%%%%%%%%%%%%%%%%%%%%%%%
%Title of paper
\title{Order-disorder transition in active nematic: A lattice model study}

\author{Rakesh Das}
\email[]{rakesh.das@bose.res.in}
\affiliation{ S N Bose National Centre for Basic Sciences, Block JD, Sector III, Salt Lake, Kolkata 700106, India} 
\author{Manoranjan Kumar} 
\email[]{manoranjan.kumar@bose.res.in}
\affiliation{ S N Bose National Centre for Basic Sciences, Block JD, Sector III, Salt Lake, Kolkata 700106, India} 
\author{Shradha Mishra}
\email[]{smishra.phy@itbhu.ac.in}
\affiliation{ S N Bose National Centre for Basic Sciences, Block JD, Sector III, Salt Lake, Kolkata 700106, India} 
\affiliation{ Department of Physics, Indian Institute of Technology (BHU), Varanasi 221005, India} 

%\date{\today}

%%%%%%%%%%%%%%%%%%%%%%%%%%%%%%%%%%%%%%%%%%%%%%%%%%%%%%%%%%%%%%%%%%%%%%%%%%%%%%%%%%%%%%%%%%%%%%%%%%%%%%%%%%%%%%%%%%%%%%%%%%%%%%%%

\begin{abstract}
We introduce a lattice model for active nematic composed of self-propelled apolar particles, 
study its different ordering states in the density-temperature parameter space, and 
compare with the corresponding equilibrium model. The active particles interact with their neighbours within 
the framework of the Lebwohl-Lasher model, and move anisotropically along their orientation to an unoccupied 
nearest neighbour lattice site. An interplay of the activity, thermal fluctuations and density 
gives rise distinct states in the system. For a fixed temperature, the active nematic shows a disordered isotropic state, 
a locally ordered inhomogeneous mixed state, and bistability between the inhomogeneous mixed and a homogeneous 
globally ordered state in different density regime. In the low temperature regime, the isotropic to the inhomogeneous 
mixed state transition occurs with a jump in the order parameter at a density less than the corresponding equilibrium 
disorder-order transition density. Our analytical calculations justify the shift in the transition density 
and the jump in the order parameter. We construct the phase diagram of the active nematic in the density-temperature 
plane.

\begin{centering}
(Received on 26 January, 2017 and accepted on 27 June, 2017 in Scientific Reports)
\end{centering}
\end{abstract}

\maketitle

%%%%%%%%%%%%%%%%%%%%%%%%%%%%%%%%%%%%%%%%%%%%%%%%%%%%%%%%%%%%%%%%%%%%%%%%%%%%%%%%%%%%%%%%%%%%%%%%%%%%%%%%%%%%%%%%%%%%%%%%%%%%%%%%

\section*{Introduction}
\noindent Self-propelled particles compose an interesting type of the active systems \cite{sriramrmp, tonertusr, rev, revvicsek, revcates}
where each particle extracts energy from its surroundings and dissipates it through motion and  collision. 
Their examples range from very small intracellular scale to larger scales 
\cite{harada, badoual, nedelec, rauch, benjacob, animalgroups, helbingnat, helbingprl, feder, feare, kuusela31, hubbard, naturevschaller, natureysumino, prlperuani2012}. 
Also many artificially designed systems, e.g., vibrated granular media \cite{vnarayanjstat, vnarayanscince, kudrolli, ncommsriram}, 
active polar disks \cite{polardisks}, active colloids \cite{activecolloid1, activecolloid2, activecolloidantoine, activecolloidTheurkauff}  
imitate the physics of the active systems. 
If ${\bf \hat{n}}$ is the average alignment direction of a collection of such active particles, and the system 
remains invariant under the transformation ${\bf \hat{n}} \rightarrow -{\bf \hat{n}}$, it is called `active nematic'. 
Activity introduces many interesting properties which are absent in their thermal equilibrium counterparts.
One of such interesting features is the presence of large density fluctuation in the ordered active nematic \cite{sradititoner}. 
Density is a  key control parameter in various experiments and numerical simulations. Earlier studies on equilibrium nematic 
for a fixed temperature show the isotropic to nematic transition at some critical density \cite{pgdgenne}. 
However, the effect of the density fluctuation in the active nematic is not well understood.

Most of the previous studies of the active systems are done either by using the coarse-grained hydrodynamic equations 
of motion \cite{shradhanjop} or microscopic rule based numerical simulation of agent based point particles \cite{ngo} 
or Brownian dynamic simulation \cite{shimanatcomm}. Here we introduce a lattice model for a two-dimensional active nematic, explore 
various states of the system in the density-temperature plane, and compare it with the corresponding equilibrium model. 
In general, lattice model itself is interesting for development of simplified theories, and provides insight into 
complex systems. Our model is analogous to the previous lattice model of polar active spins \cite{solonprl, solonpre}; but we include 
volume exclusion to avoid multiple occupancy on single site. Such volume exclusion limits the motion of particles towards 
an occupied neighbouring site, and introduces new features, e.g., typical pattern formation \cite{prlperuani2012, prlperuani2011}, 
density induced motility \cite{dimprl} in the system.

\begin{figure*}[t]
  \includegraphics[width=\linewidth]{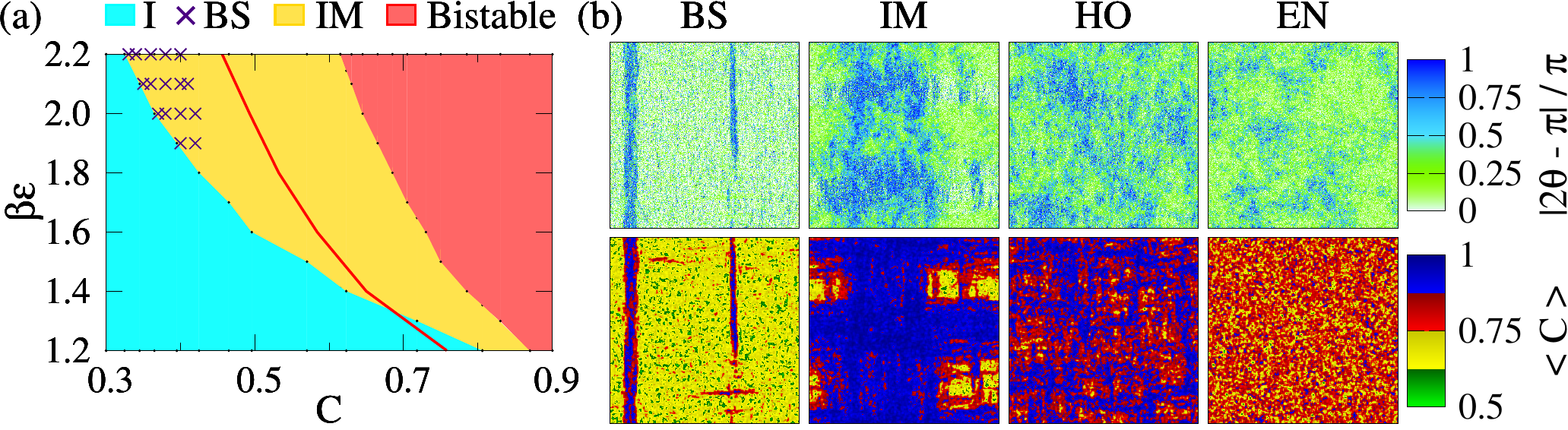}
  \caption{ Phase diagram.
  (a) Phase diagram for both the equilibrium and the active nematic in the density - inverse temperature plane. 
  The equilibrium system remains in the isotropic (EI) state in the low density regime (on the left of the solid line) 
  and in the nematic (EN) state in the high density regime (on the right of the solid line). 
  The active nematic goes from the disordered isotropic (I) state to the locally ordered inhomogeneous mixed (IM) state 
  with increasing density or decreasing temperature. The I - IM transition occurs with the appearance of clear bands 
  (BS) in the low temperature regime. In the high density regime the active nematic shows bistability between the IM and 
  the homogeneous globally ordered (HO) state.
  (b) Upper panel shows particle inclination towards the horizontal direction. Colour bar ranging from zero to one indicates 
  vertical to horizontal orientation, respectively. BS is the banded state configuration shown for $(\beta\epsilon, C)=(2.0, 0.38)$. 
  IM, HO and EN state configurations are shown for $(\beta\epsilon, C)=(2.0, 0.78)$. 
  Lower panel shows the coarse-grained density in the respective states. }
\label{fig:phase_snap}
\end{figure*}

\begin{figure}[t]
  \centering
  \includegraphics[width=0.6\linewidth]{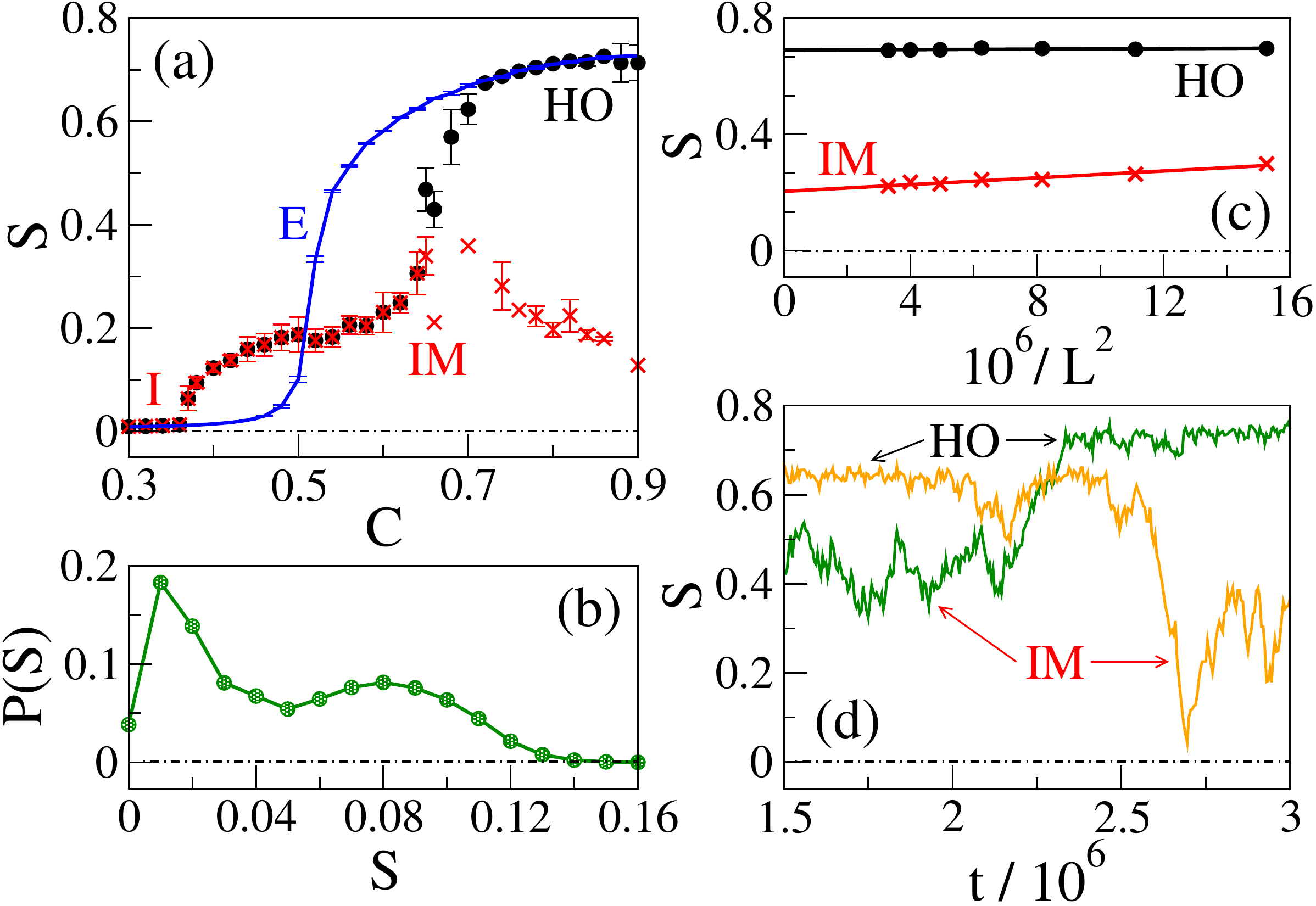}
  \caption{ Disorder - order transition.
  (a) Scalar order parameter versus packing density plot for $400$ $\times$ $400$ system size at $\beta\epsilon=2.0$.
  The equilibrium system (E, solid line) shows continuous isotropic to nematic state transition with increasing density.
  The active system goes from the isotropic (I) state to the locally ordered inhomogeneous mixed (IM, $\times$) state. 
  In the high density regime the system shows bistability between the IM state and the homogeneous globally ordered (HO, $\bullet$) state.
  (b) The I to IM transition at low temperature occurs with a jump in $S$ where the particles form bands (BS). Distribution of 
  the scalar order parameter near the I - BS transition at $(\beta\epsilon, C)=(2.0,0.37)$ shows two peaks. 
  (c) Finite size scaling of $S$ for both the HO and the IM state at $(\beta\epsilon, C)=(2.0,0.76)$. 
  (d) Order parameter time series show that the active system flips in between the HO and the IM state in the bistable regime.
      Two time series are shown for two different parameter values in the high density regime.}
\label{fig:S_C_diag}
\end{figure}

We construct a phase diagram for the active nematic in the density-temperature plane, as shown in Fig. \ref{fig:phase_snap}(a). 
There we observe - (i) disordered isotropic (I) state in low density regime, (ii) locally ordered inhomogeneous mixed (IM) state 
in intermediate density regime, and (iii) bistability between the IM and a homogeneous globally ordered (HO) state in high density 
regime. In contrast to the continuous isotropic to nematic (I-N) transition in the equilibrium system, the I to IM state transition 
in the active nematic in the low temperature regime occurs with a jump in order parameter, as shown in Fig. \ref{fig:S_C_diag}(a). 
This transition occurs at a density lower than the equilibrium critical value, and the system forms clear bands (BS) in this regime. 
We finally justify the jump in the order parameter and the shift in the transition density by analytical study of the coarse-grained 
hydrodynamic equations written for the active model.

%%%%%%%%%%%%%%%%%%%%%%%%%%%%%%%%%%%%%%%%%%%%%%%%%%%%%%%%%%%%%%%%%%%%%%%%%%%%%%%%%%%%%%%%%%%%%%%%%%%%%%%%%%%%%%%%%%%%%%%%%%%%%%%%

%%%%%%%%%%%%%%%%%%%%%%%%%%%%%%%%%%%%%%%%%%%%%%%%%%%%%%%%%%%%%%%%%%%%%%%%%%%%%%%%%%%%%%%%%%%%%%%%%%%%%%%%%%%%%%%%%%%%%%%%%%%%%%%%%

%%%%%%%%%%%%%%%%%%%%%%%%%%%%%%%%%%%%%%%%%%%%%%%%%%%%%%%%%%%%%%%%%%%%%%%%%%%%%%%%%%%%%%%%%%%%%%%%%%%%%%%%%%%%%%%%%%%%%%%%%%%%%%%%

\begin{figure}[t]
  \centering
  \includegraphics[width=1.0\linewidth]{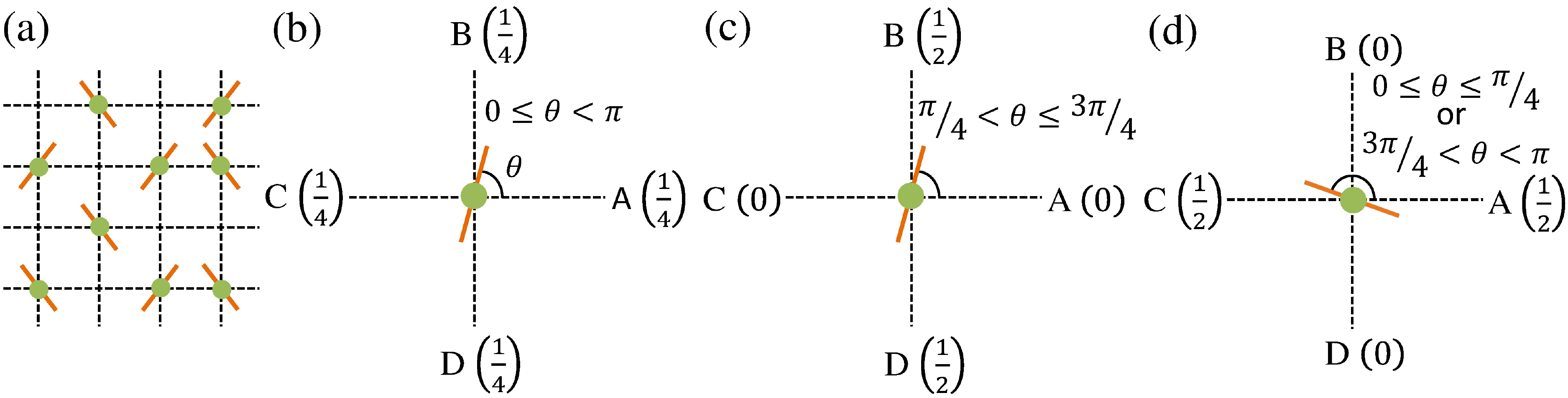}
  \caption{ Model figure.
  (a) Two dimensional square lattice with occupied ($n=1$) or unoccupied ($n=0$) sites. Filled circles 
  indicate the occupied sites. Inclinations of the rods towards the horizontal direction show respective 
  particle orientations $\theta\in\left[0,\pi\right]$. (b) Equilibrium move: particle can move to any of the four 
  neighbouring sites with equal probability $1/4$. (c, d) Active move: particle can move to either of 
  its two neighbouring sites with probability $1/2$, if unoccupied, in the direction it  is more inclined to, 
  i.e., along $BD$ in (c), and $AC$ in (d). }
\label{fig:model}
\end{figure}

%%%%%%%%%%%%%%%%%%%%%%%%%%%%%%%%%%%%%%%%%%%%%%%%%%%%%%%%%%%%%%%%%%%%%%%%%%%%%%%%%%%%%%%%%%%%%%%%%%%%%%%%%%%%%%%%%%%%%%%%%%%%%%%%%%%%%%%%%%%%%%%%%%

\section*{Model}\label{secmodel}
\noindent  We consider a collection of apolar particles on a two dimensional square lattice, as shown in schematic 
diagram Fig. \ref{fig:model}(a). Occupation number `$n_i$' of the $i^{th}$ lattice site can take values $1$ (occupied) 
or $0$ (unoccupied). Orientation $\theta_i$ of apolar particle at the $i^{th}$ site can take any value between $0$ and $\pi$. 
The model follows two sequential processes at every step; first, a particle moves to a nearest neighbouring site with 
{\it some probability}, and then orientation of the particle is updated based on its nematic interaction with its nearest 
neighbours. We define two kinds of models on the basis of particle movement: (i) `Equilibrium model' (EM) - particle moves 
with equal probability $1/4$ to any of the four neighbouring sites (Fig. \ref{fig:model}(b)), (ii) `Active model' (AM) - 
in this model particle movement occurs in two steps. First, it chooses a direction along which it is more inclined. 
As shown in Fig. \ref{fig:model}(c,d), it chooses the direction of movement along $BD$ if $ \pi /4 < \theta \le 3\pi/4$ and 
along $AC$ otherwise. In the second step, it moves to a randomly selected site between the two nearest neighbouring sites 
along the chosen direction. For example, if $BD$ is selected as the direction of movement, then the particle moves to 
randomly selected site $B$ or $D$ in the second step. In both the models, we consider volume exclusion, i.e., particle 
movement is allowed only if the selected site is unoccupied. 
  
In both the models, the particles also interact with their nearest neighbours. The interaction depends on the relative 
orientation of the particles and is represented by a modified Lebwohl-Lasher Hamiltonian \cite{llasher}
\begin{equation}
\mathcal{H} = -\epsilon \sum_{<ij>}n_i n_j \cos2(\theta_i-\theta_j)
\label{eqll}
\end{equation}
where $\epsilon$ is the interaction strength between two neighbouring particles. The interaction in equation (\ref{eqll}) 
governs the orientation update of the particle. We employ Metropolis Monte-Carlo (MC) algorithm \cite{mcbinder} for 
orientation update of the particle after the movement trial. In both the models, an order parameter defining the global
alignment of the system does not remain conserved during the MC orientation update described above. In actual granular 
or biological systems where mutual alignment emerges because of steric repulsion, orientation of particles need not 
to follow a conservation law. An order parameter defined by coarse-graining the orientation in our present model is 
a class of non-conserved order parameter: {\it Model A} as described by Hohenberg and Halperin \cite{HohenbergHalperin}.

Both the models EM and AM comprise of two different physical aspects - motion of the particles and nematic interaction 
amongst the nearest neighbours. If the particles are not allowed to move, the models reduce to an apolar analogue of the 
diluted XY-model with nonmagnetic impurities \cite{dilutedxymodel}, where impurities and spins are analogous to vacancies 
and particles, respectively. However, unlike the diluted XY-model, particles in these models are dynamic. In the EM, the 
particle diffuses to neighbouring sites, whereas it moves anisotropically in the AM. The anisotropic movement of the active 
particles arises in general because of the self-propelled nature of the particles in many biological \cite{kemkemer} and 
granular systems \cite{vnarayanjstat, vnarayanscince}. This move produces an active curvature coupling current in the coarse-grained hydrodynamic 
equations of motion \cite{shradhanjop, sradititoner}. The AM does not satisfy the detailed balance principle \cite{mcbinder}, 
because of the orientation update after the anisotropic movement. The coupling of the particle movement with the orientation 
update in our active model is analogous to the active Ising spin model introduced by Solon and Tailleur \cite{solonprl, solonpre}, where 
the probabilistic flip of the spins is an equilibrium process, whereas the out-of-equilibrium aspect of the model is attributed 
to the anisotropic movement probability of the spins. However, their orientation update algorithm \cite{solonprl, solonpre} is similar 
to kinetic Monte-Carlo, whereas we use Metropolis Monte-Carlo algorithm to update particle orientation.

%%%%%%%%%%%%%%%%%%%%%%%%%%%%%%%%%%%%%%%%%%%%%%%%%%%%%%%%%%%%%%%%%%%%%%%%%%%%%%%%%%%%%%%%%%%%%%%%%%%%%%%%%%%%%%%%%%%%%%%%%%%%%%%%%%%%%%%%%%%%%%%%%%

\section*{Numerical Study}\label{secnumerical}
\noindent We consider a collection of $N$ particles with random orientation $\theta \in [0,\pi]$ homogeneously distributed 
on a $L \times L$ lattice ($L=256, 400, 512$) with periodic boundary. Packing density of the system is defined as 
$C=N/(L \times L)$. We choose a particle randomly, move it to a neighbouring site obeying exclusion, and then update its 
orientation using Metropolis algorithm. In each iteration, we repeat the same process for $N$ number of times, and we use 
$1.5\times10^6$ iterations to achieve the steady state of the system. We obtain the steady state results by averaging 
the observables over next $1.5 \times 10^6$ iterations and use more than twenty realisations for better statistics. 

The ordering in the system is characterised by a scalar order parameter defined as
\begin{equation}
S=\sqrt{\left(\frac{1}{N}\sum_i n_i \cos(2 \theta_i)\right)^2+\left(\frac{1}{N}\sum_i n_i \sin(2 \theta_i)\right)^2}.
\label{eqops}
\end{equation}
It is proportional to the positive eigenvalue of the nematic order parameter $\Q$ \cite{pgdgenne}. It takes the minimum 
value $0$ in the disordered state and the maximum value $1$ in the complete ordered state. First we study the EM as a 
function of inverse temperature $\beta= 1 / k_BT$ for different packing densities. As shown in Fig. \ref{figtemp},
the system shows disordered isotropic to nematic state (I-N) transition with decreasing temperature. In contrast to the 
first order I-N transition in the equilibrium Lebwohl-Lasher model in three dimensions \cite{llasher,chaiklub}, we find 
continuous transition for the EM defined in two dimensions. The observed nature of transition supports the study by 
Mondal and Roy \cite{mondalroy}. Similar to the diluted XY-model \cite{dilutedxymodel}, the critical inverse temperature 
$\beta_c(C)$ increases with density in the EM. 

%%%%%%%%%%%%%%%%%%%%%%%%%%%%%%%%%%%%%%%%%%%%%%%%%%%%%%%%%%%%%%%%%%%%%%%%%%%%%%%%%%%%%%%%%%%%%%%%%%%%%%%%%%%%%%%%%%%%%%%%%%%%%%%%%%%%%%%%%%%%%%%%%%

\subsection*{Phase diagram}
\noindent We construct phase diagram for both the equilibrium model and the active model on the density-temperature 
plane. As shown in Fig. \ref{fig:phase_snap}(a), two distinct states appear in the EM - (i) an equilibrium
isotropic (EI) state on the left side of the red boundary and (ii) an equilibrium nematic (EN) state on the
right side of the red boundary. In the EI state, particles 
remain disordered and homogeneously distributed throughout the system. Consequently, the scalar order parameter
$S\simeq0$ in this state. With increasing density or decreasing temperature the particles get mutually ordered 
and form the EN state ($S>0$). As shown in Fig. \ref{fig:S_C_diag}(a), for a fixed temperature the scalar
order parameter increases continuously with increasing density, and the system enters into the nematic state.
Both the particle orientation and the coarse-grained density remain homogeneous in the EN state, as shown in the 
real space snapshot Fig. \ref{fig:phase_snap}(b).

Similar to the EM, the active system remains in a homogeneous disordered isotropic (I) state 
in the high temperature and/or low packing density regime (cyan coloured regime in the phase 
diagram Fig. \ref{fig:phase_snap}(a)). 
With increasing density or decreasing temperature, beyond the I state, the active system enters into an inhomogeneous 
mixed (IM) state (golden regime in the phase diagram Fig. \ref{fig:phase_snap}(a)), where locally ordered high-density 
domains coexist with disordered low-density regions. In the low temperature regime ($\beta\epsilon \in [1.9, 2.2]$),
the I to IM state transition with increasing $C$ occurs with a jump in the scalar 
order parameter $S$, as shown in Fig. \ref{fig:S_C_diag}(a). In the very beginning of the IM state, as indicated by 
cross symbols in Fig. \ref{fig:phase_snap}(a), we find a banded state (BS) in the low temperature regime, where
particles cluster and align themselves within a strip to form band. However, out of the strip
the system remains disordered with low local density, as shown in the real space snapshot Fig. \ref{fig:phase_snap}(b). 
On further increment of the packing density $C$,
bands formed in different directions start mixing leaving the system with many locally ordered high density patches 
separated by low density disordered regions. Typical real space snapshots for the orientation and the coarse grained density in 
the IM state are shown in Fig. \ref{fig:phase_snap}(b). The jump in the $S - C$ curve reduces with increasing
temperature, and no bands appear in the high temperature ($\beta\epsilon < 1.9$) regime. 

%%%%%%%%%%%%%%%%%%%%%%%%%%%%%%%%%%%%%%%%%%%%%%%%%%%%%%%%%%%%%%%%%%%%%%%%%%%%%%%%%%%%%%%%%%%%%%%%%%%%%%%%%%%%%%%%%%%%%%%%%%%%%%%%

\begin{figure}[b]
  \centering
  \includegraphics[width=0.4\linewidth]{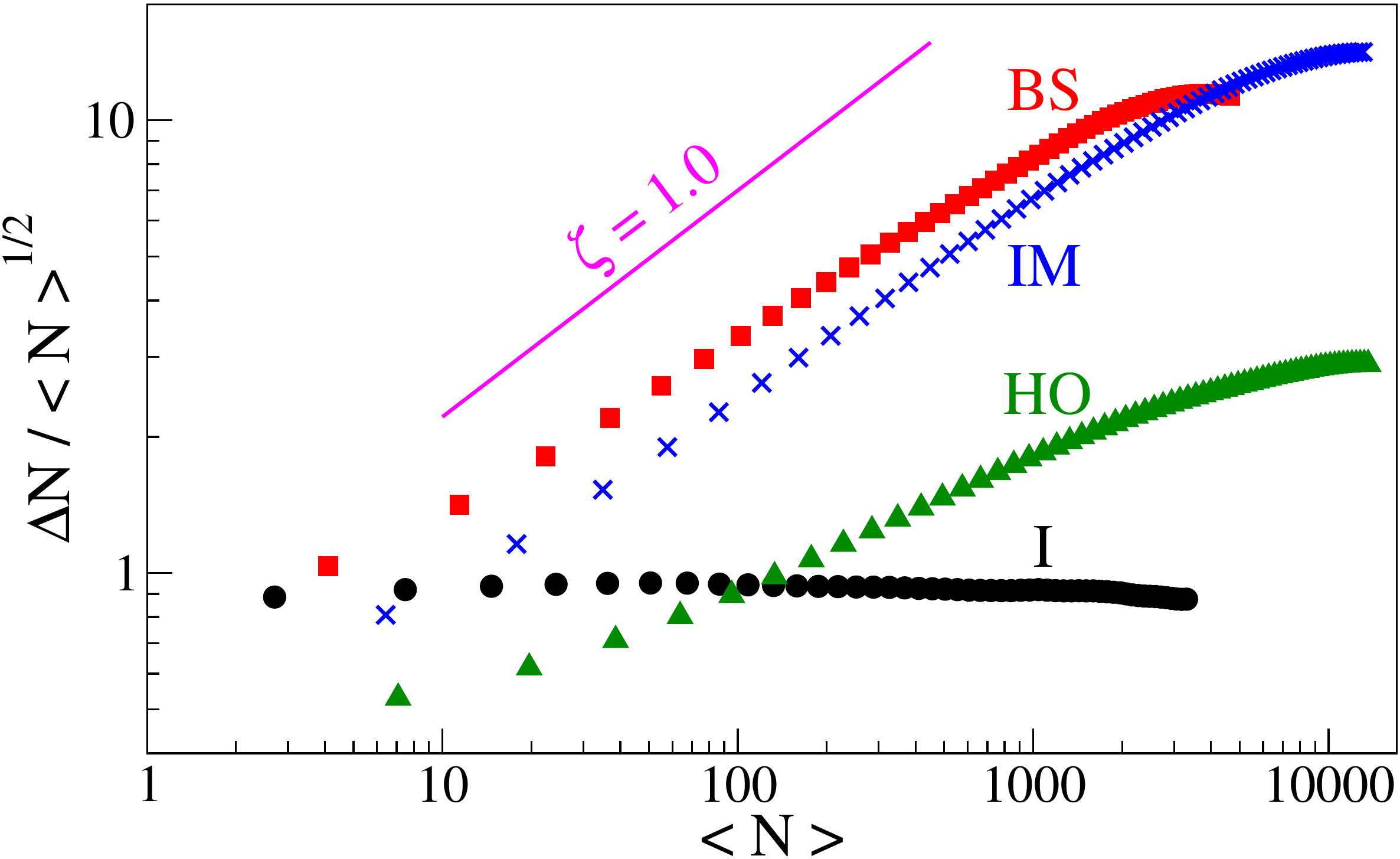}
  \caption{ Density fluctuation $\Delta N = \sqrt{<N^2> - <N>^2}$.
  All the active ordered states show large density fluctuation obeying the relation $\Delta N \sim \left<N\right>^{\zeta}$
  with $\zeta > 1/2$. The active disordered isotropic state shows normal density fluctuation with $\zeta = 1/2$.}
  \label{fig:gnf}
\end{figure}

%%%%%%%%%%%%%%%%%%%%%%%%%%%%%%%%%%%%%%%%%%%%%%%%%%%%%%%%%%%%%%%%%%%%%%%%%%%%%%%%%%%%%%%%%%%%%%%%%%%%%%%%%%%%%%%%%%%%%%%%%%%%%%%%

Figure \ref{fig:S_C_diag}(a) shows that the I to BS transition occurs in the low temperature regime with a jump in $S$
at a density lower than the corresponding equilibrium I-N transition density $C_{IN}$. These bands appear because of the large activity strength. 
A linear stability analysis, as detailed later in this paper, shows that the large activity strength 
induces an instability in the disordered isotropic state. This instability goes away for small activity strength or at high
temperature. We also do a renormalised mean field calculation of an effective free energy written for the active 
nematic. The calculation predicts a jump in the scalar order parameter and shows a shift in the disordered 
$(S=0)$ to ordered $(S\ne 0)$ state transition density. Both the jump in $S$ and the shift in the transition density 
reduce with the activity strength or increasing temperature. The I to BS transition is a first order transition.
The shift in the disorder-order transition point is a common feature of the active systems. 
For large activity and low temperature, if the system density is above a certain value but less than $C_{IN}$, 
the large density fluctuation present in these systems causes local alignment with local density higher than $C_{IN}$. 
Large density fluctuation is an intrinsic feature of the active systems, 
and as shown in Fig. \ref{fig:gnf}, we also observe the same in the ordered active states in our model.
Due to activity these locally ordered regions move anisotropically and 
combine with nearby region with similar local ordering. So larger ordered region forms at 
mean density lower than the equilibrium I-N transition density. 
Therefore, we find a disordered to ordered state transition at a lower density.
For large activity strength, I-BS transition occurs with the jump in scalar order parameter.
In our numerical study we calculate the probability $P(S)$ of the scalar order parameter 
averaging over many iterations and realisations near the I-BS transition point. Figure \ref{fig:S_C_diag}(b) shows 
$P(S)$ has two peaks, which further supports the first order I-BS transition for large activity strength.

In the high density regime (red coloured regime in the phase diagram Fig. \ref{fig:phase_snap}(a)), the AM shows bistability,
i.e., it can be either in the locally ordered IM state or in a homogeneous globally ordered (HO) state. 
As shown in Fig. \ref{fig:S_C_diag}(a), the $S - C$ curve for fixed temperature bifurcates in the high density 
regime; the lower branch corresponds to the earlier discussed IM state, whereas the higher branch indicates
the existence of the globally ordered state. Figure \ref{fig:phase_snap}(b) shows that the system possesses less density 
inhomogeneity in the HO state compared to the IM state. A finite size scaling of both the HO and the IM state, as shown in the Fig. 
\ref{fig:S_C_diag}(c), shows that the active nematic possesses non-zero finite order in both these states. 
Order parameter time series shown in Fig. \ref{fig:S_C_diag}(d) confirms the bistability of the system in the high density regime.
Bistability is not generally seen in other agent based numerical simulations of point particles \cite{ngo};
it appears because of finite filling constraint of the model. This feature can be suppressed if we allow more 
than one particle to sit together. In the complete filling limit $C =1.0$, the AM is equivalent to the EM, 
and it shows the globally ordered HO state only.

%%%%%%%%%%%%%%%%%%%%%%%%%%%%%%%%%%%%%%%%%%%%%%%%%%%%%%%%%%%%%%%%%%%%%%%%%%%%%%%%%%%%%%%%%%%%%%%%%%%%%%%%%%%%%%%%%%%%%%%%%%%%%%%%

\begin{figure}[b]
  \includegraphics[width=0.6\linewidth]{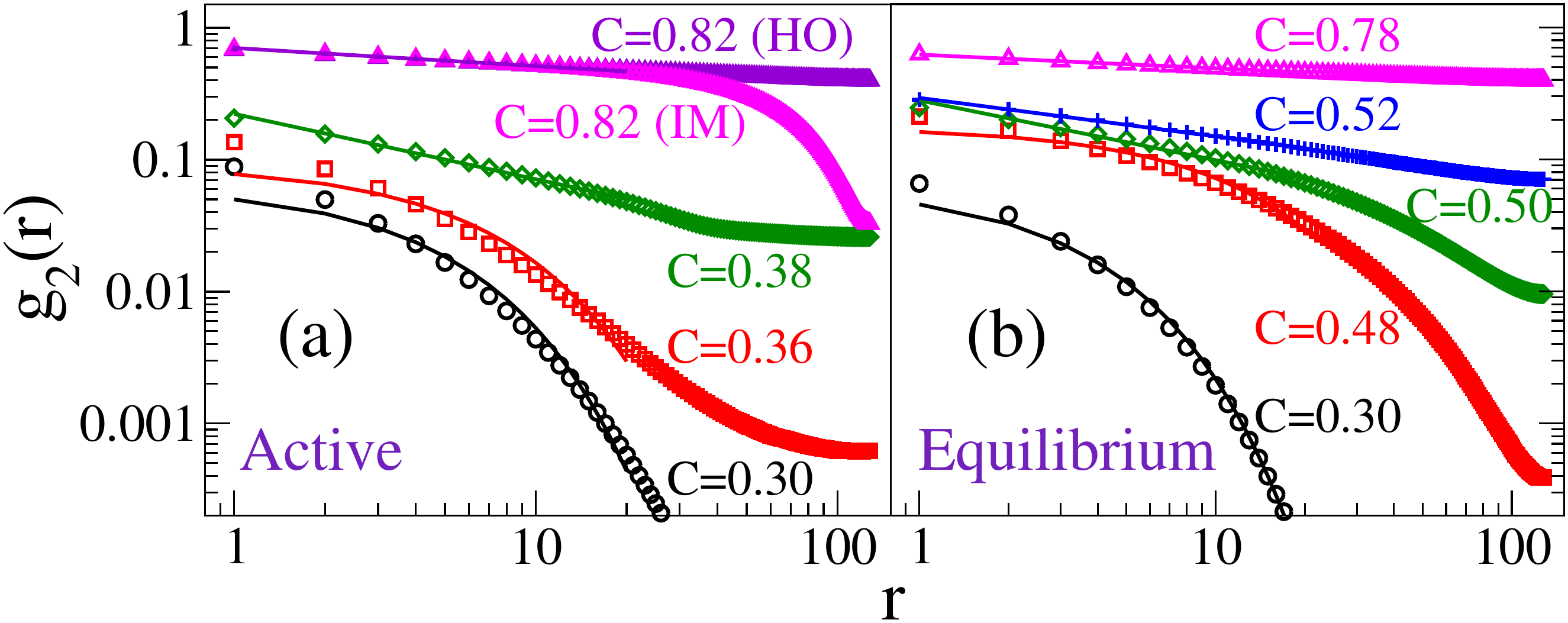}
  \caption{ Two-point orientation correlation shown for $\beta\epsilon=2.0$ on log-log scale.
  (a) Active model: $g_2(r)$ decays exponentially at low density ($\bigcirc$, $\square$) and algebraically at
  high density ($\diamond$, $\bigtriangleup$). In the bistable regime at high density ($\bigtriangleup$), $g_2(r)$ 
  decays algebraically in the HO state and abruptly in the IM state.
  (b) Equilibrium model: $g_2(r)$ decays exponentially at low density ($\bigcirc$, $\square$) and algebraically
  at high density ($\diamond$, $+$, $\bigtriangleup$). Continuous lines are the respective fits, fitted for more 
  than one decade.}
  \label{fig:correlation}
\end{figure}

%%%%%%%%%%%%%%%%%%%%%%%%%%%%%%%%%%%%%%%%%%%%%%%%%%%%%%%%%%%%%%%%%%%%%%%%%%%%%%%%%%%%%%%%%%%%%%%%%%%%%%%%%%%%%%%%%%%%%%%%%%%%%%%%%%%%%%%%%%%%%%%%%%

\subsection*{Two-point orientation correlation}
\noindent We further characterise various states on the basis of the two-point orientation correlation 
in the different states of the equilibrium and the active nematic. It is defined as
$ g_2(r) = <\sum_i n_i n_{i+r} $ $\cos\left[2\left(\theta_i-\theta_{i+r}\right)\right]/\sum_i n_i^2 > $ 
where $r$ represents interparticle distance, and $< . >$ signifies an average over many realisations. 
Figure \ref{fig:correlation}(a, b) show $g_2(r)$ versus $r$ plots on log-log scale for the AM and the EM,
respectively, for a fixed inverse temperature $\beta\epsilon=2.0$. In the AM, $g_2(r)$ decays exponentially at low
packing density $C<0.37$, i.e., in the isotropic state. Therefore, the active isotropic is a short-range-ordered (SRO) state.
In the BS at $C=0.38$, $g_2(r)$ decays following a power law. Therefore, the system is in a quasi-long-range-ordered 
(QLRO) state. Ordering increases with density. 
At high packing density, correlation function confirms the bistability in the active system.
At $C=0.82$, $g_2(r)$ shows power law decay in the HO state, whereas in the IM state $g_2(r)$ decays abruptly 
after a distance $r$. The abrupt change in $g_2(r)$ at a certain distance indicates the presence of locally 
ordered clusters in the IM state.
In contrast, the equilibrium system shows a transition from SRO (exponential decay) 
isotropic state at low density $C\lsim0.48$ to QLRO (power law decay) nematic state at high density $C\gsim0.50$. 

%%%%%%%%%%%%%%%%%%%%%%%%%%%%%%%%%%%%%%%%%%%%%%%%%%%%%%%%%%%%%%%%%%%%%%%%%%%%%%%%%%%%%%%%%%%%%%%%%%%%%%%%%%%%%%%%%%%%%%%%%%%%%%%%

\begin{figure}[b]
  \includegraphics[width=0.5\linewidth]{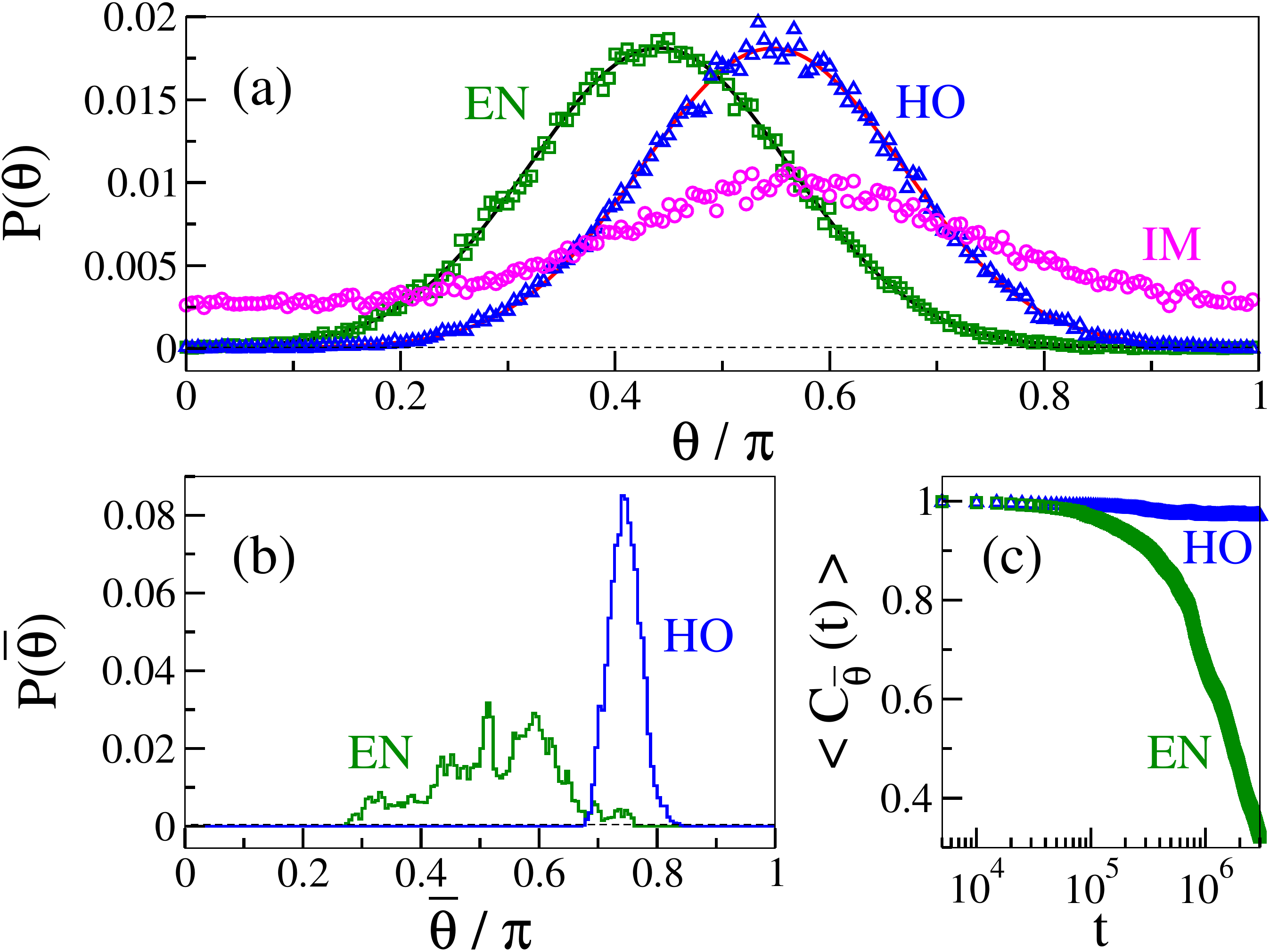}
  \caption{Steady state characteristics of high density states. 
  (a) Orientation distribution $P(\theta)$ of particles calculated from one snapshot in the steady state. 
  $P(\theta)$ fits with Gaussian distribution (continuous lines) for both the HO and the EN states. 
  The IM state shows broad distribution of $\theta$. 
  (b) Distribution of the mean orientation $P(\bar \theta)$ calculated from $\bar \theta$ of each snapshot 
  in the steady state. $P(\bar \theta)$ is broad for the EN state in comparison to the HO state. 
  (c) Steady state autocorrelation $C_{\bar \theta}(t)$ of the mean orientation of the system. 
  All plots are shown for $(\beta\epsilon, C)=(2.0,0.80)$. }
  \label{fig:distribution}
\end{figure}

%%%%%%%%%%%%%%%%%%%%%%%%%%%%%%%%%%%%%%%%%%%%%%%%%%%%%%%%%%%%%%%%%%%%%%%%%%%%%%%%%%%%%%%%%%%%%%%%%%%%%%%%%%%%%%%%%%%%%%%%%%%%%%%%%%%%%%%%%%%%%%%%%%

\subsection*{Orientation distribution  and autocorrelation of the mean orientation}
\noindent We compare  the steady state properties of the active and the equilibrium models in the high density limit. 
First we calculate the steady state (static) orientation distribution $P(\theta)$ from a snapshot of particle 
orientation $\theta$. As shown in Fig. \ref{fig:distribution}(a), both the active HO and the equilibrium nematic show 
Gaussian distribution of orientation. Peak position of $P(\theta)$ for both the EN and the HO state can appear at any 
point between $0$ and $\pi$ because of the continuous broken rotational symmetry of the Hamiltonian shown in equation 
(\ref{eqll}). Data shown in Fig. \ref{fig:distribution}(a) is for one realisation only, and for other realisations also
the distribution $P(\theta)$ remains Gaussian with peak at other $\theta$ values. Therefore, orientation fluctuation of 
the particles in the active HO state is same as in the equilibrium nematic state. The distribution $P(\theta)$ in the IM 
state is very broad and spans over the whole range of orientation. Therefore, the system possess no global ordering in 
the IM state. 

We also calculate the time averaged distribution $P(\bar \theta)$ of mean orientation of all the particles in the active 
HO and the equilibrium nematic states. The mean orientation $\bar \theta(t)$ of all particles is calculated for each 
iteration time $t$ in the steady state. The distribution $P(\bar \theta)$ of the mean orientation is obtained from these 
$\bar \theta(t)$ data. This distribution is a measure of the fluctuation in the global orientation of the particles in 
the steady state. As shown in Fig. \ref{fig:distribution}(b), $P(\bar \theta)$ in the active HO state is narrow in 
comparison to the broad distribution in the EN state.  We also calculate the autocorrelation of the mean orientation
$C_{\bar\theta}(t)=<\frac{1}{t} \sum_{\tau=1}^t cos\left[ 2\left\{ \bar\theta\left(t_0\right)-\bar\theta\left(t_0+\tau\right) \right\} \right]>$
in the steady state. As shown in Fig. \ref{fig:distribution}(c), $C_{\bar \theta}(t)$ decreases with time in the EN state, 
but remains unchanged in the active HO state. Both these results imply that the fluctuation in the global orientation 
direction $\bar \theta$ in the active HO state is small compared to the EN state. We do not calculate the mean orientation 
$\bar \theta$ in the active IM state, because the system possesses no global ordering in this state.

%%%%%%%%%%%%%%%%%%%%%%%%%%%%%%%%%%%%%%%%%%%%%%%%%%%%%%%%%%%%%%%%%%%%%%%%%%%%%%%%%%%%%%%%%%%%%%%%%%%%%%%%%%%%%%%%%%%%%%%%%%%%%%%%

\section*{Phenomenological approach to understand low density states of the active model}\label{sechydro}
\noindent In this section we write the hydrodynamic equations of motion for the active model and characterise
the low density states of the system. The equations of motion for the slow variables of the system, 
i.e., the number density $\wp({\bf r}, t) = \sum_{l}\delta({\bf r}-{\bf R}_l(t))$ and the order parameter
$ {\Pi}_{i j}({\bf r}, t)  = \wp({\bf r}, t) \Q_{ij}({\bf r}, t) = \sum_{l} 
({\bf m}_{li} {\bf m}_{lj} - \frac{1}{2}\delta_{i j}) \delta({\bf r}-{\bf R}_l(t))$  
are as follows \cite{shradhanjop, sradititoner}:
\begin{equation}
\partial_{t}\wp=a_{0}\nabla_{i}\nabla_{j}{\Pi}_{ij}+D_{\wp}\nabla^{2}\wp
\label{eqdensity}
\end{equation}
and
\begin{align}
\partial_{t}{\Pi}_{ij} =\left\{ \alpha_{1}\left(\wp\right) - \alpha_{2}\left(\Pi:\Pi\right)\right\} \Pi_{ij} 
 + \beta\left(\nabla_{i}\nabla_{j}-\frac{1}{2}\delta_{ij}\nabla^{2}\right)\wp+D_{\Pi}\nabla^{2}{\Pi}_{ij}
\label{eqop}
\end{align}
Here ${\bf R}_l(t)$ represents position of the particle $l$, and ${\bf m}_l = (\cos \theta_l, \sin \theta_l)$
is the unit vector along the orientation $\theta_l$.
The total number of particles being a conserved quantity of the system, equation (\ref{eqdensity}) represents a continuity 
equation $\partial_{t}\wp = -\nabla \cdot {\bf J}$ where the current $J_i = -a_0 \nabla_j {\Pi}_{ij} - D_{\wp} \nabla_i \wp$.
The first term of $J_i$ consists of two parts: an anisotropic diffusion current ${\bf J}_{p1} \propto \Q_{ij}\nabla_i \wp$
and an active curvature coupling current ${\bf J}_a \propto a_0 {\wp \nabla_j \Q_{ij}}$ where $a_0$ is the activity strength 
of the system. The second term represents an isotropic diffusion  ${\bf J}_{p2} \propto \nabla \wp$. The $\alpha$ terms in 
equation (\ref{eqop}) represent mean field alignment in the system. We choose $\alpha_1(\wp) = (\frac{\wp}{\wp_{IN}}-1)$ as
a function of density that changes sign at some critical density $\wp_{IN}$. The $\beta$ term represents coupling with density. 
The last term represents diffusion in order parameter that is written under equal elastic constant approximation for two-dimensional 
nematic. The steady state solution $\wp({\bf r}, t) = \wp_0$ and ${\Pi}({\bf r}, t) = {\Pi}_{0}$, where 
${\Pi}_0 = \sqrt{\frac{\alpha_1\left(\wp_0\right)}{\alpha_2}}$, of equations (\ref{eqdensity}) and (\ref{eqop}) represents a 
homogeneous ordered state for $\alpha_1(\wp_0) >0 $ at $\wp_0 > \wp_{IN}$, and a disordered isotropic state for 
$\alpha_1(\wp_0) < 0$ at $\wp_0 < \wp_{IN}$. 

%%%%%%%%%%%%%%%%%%%%%%%%%%%%%%%%%%%%%%%%%%%%%%%%%%%%%%%%%%%%%%%%%%%%%%%%%%%%%%%%%%%%%%%%%%%%%%%%%%%%%%%%%%%%%%%%%%%%%%%%%%%%%%%%

%\subsection{Linear stability of disordered isotropic state}\label{linearised}
We study the linear stability of the disordered isotropic state ({\bf $\Pi_0 = 0$}) by examining the dynamics of spatially 
inhomogeneous fluctuations $\delta \wp({\bf r}, t) = \wp({\bf r}, t) - \wp_0$, $\delta {\Pi}_{11} = {\Pi}_{11}({\bf r}, t)$, 
and $\delta {\Pi}_{12} = {\Pi}_{12}({\bf r}, t)$. We obtain the linearised coupled equations of motion for small fluctuations as
\begin{eqnarray}
\partial_t \delta \wp &=& a_0 \left( \partial_x^2 - \partial_y^2 \right) \delta {\Pi}_{11}
                         + 2a_0\partial_x\partial_y\delta{\Pi}_{12}
                         + D_{\wp}\nabla^2\delta\wp                      \label{eq_del_rho} \\
\partial_t \delta {\Pi}_{11} &=& \alpha_1\left(\wp_0\right) \delta{\Pi}_{11} 
				+ D_{\Pi}\nabla^2\delta{\Pi}_{11}
				+ \frac{\beta}{2}\left( \partial_x^2 - \partial_y^2 \right) \delta \wp  \label{eq_del_w_11} \\
\partial_t \delta {\Pi}_{12} &=& \alpha_1\left(\wp_0\right) \delta{\Pi}_{12}
				+ D_{\Pi}\nabla^2\delta{\Pi}_{12}
				+ \beta\partial_x\partial_y\delta\wp      \label{eq_del_w_12}
\end{eqnarray}
Using Fourier transformation
\begin{equation}
Y\left({\bf q},\lambda\right) = \int e^{i{\bf q}.{\bf r}} e^{\lambda t} Y\left({\bf r},t\right) d{\bf r} dt 
\end{equation}
we get linear set of equations in the Fourier space as
\begin{eqnarray}
\lambda
\left( \begin{array} {ccc}
\delta\wp \\
\delta{\Pi}_{11} \\
\delta{\Pi}_{12}
\end{array} \right) = 
M
\left( \begin{array} {ccc}
\delta\wp \\
\delta{\Pi}_{11} \\
\delta{\Pi}_{12}
\end{array} \right)
\label{eq_linear}
\end{eqnarray}
where $M$ is the coefficient matrix as obtained from equations (\ref{eq_del_rho}), (\ref{eq_del_w_11}), and 
(\ref{eq_del_w_12}) after the transformation.
We solve equation (\ref{eq_linear}) for the hydrodynamic modes $\lambda$. We choose $q_x=q_y=\frac{q}{\sqrt{2}}$ 
since both the directions are equivalent. Therefore, we obtain 
\begin{equation}
 \left( \lambda -  \alpha_1 \left( \wp_0 \right) + D_{\Pi} q^2 \right) 
 \left\{ \left( \lambda + D_{\wp} q^2 \right) \left( \lambda - \alpha_1 \left( \wp_0 \right)  + D_{\Pi} q^2 \right) - \frac{1}{2} a_0 \beta q^4 \right\}  = 0
\end{equation}
For small wave-vector $q$, we can find an unstable mode
\begin{equation}
\lambda_{+} = -2 D_{\wp} q^2 + \frac{a_0 \beta q^4}{2 |\alpha_1(\wp_0)|} - \frac{a_0 \beta q^6(D_{\Pi}-D_{\wp})}{\alpha_1^2(\wp_0)}
\label{modeeq}
\end{equation}
For small $D_{\wp}$ and large actvitity $a_0$ this mode becomes unstable for $q < q_c$, where
\begin{equation}
q_c^2 = \frac{|\alpha_1(\wp_0)|}{2\Delta D} + \frac{1}{2}\sqrt{\left(\frac{|\alpha_1(\wp_0)|}{\Delta D}\right)^2 - \frac{8 D_{\wp} \alpha_1^2}{\Delta D a_0 \beta}}
\label{eqq}
\end{equation}
provided $\Delta D = D_{\Pi}-D_{\wp}$ is positive and $a_0 \beta > 8 D_{\wp} \Delta D$. Therefore, the unstable mode 
$\lambda_+$ causes the I - BS transition for small diffusivity, i.e., at low temperature, and for large activity strength $a_0$.

%%%%%%%%%%%%%%%%%%%%%%%%%%%%%%%%%%%%%%%%%%%%%%%%%%%%%%%%%%%%%%%%%%%%%%%%%%%%%%%%%%%%%%%%%%%%%%%%%%%%%%%%%%%%%%%%%%%%%%%%%%%%%%%%

%\subsection{Renormalised mean field study for small $S$}\label{rmf}
We also calculate the jump in the scalar order parameter $S$ and the shift  in the transition density from equations 
(\ref{eqdensity}) and (\ref{eqop}). A homogeneous steady state solution of these equations gives a mean field transition
from the isotropic to the nematic state at density $\wp_{IN}$ where $\alpha_1(\wp)$ changes sign. Using renormalised mean 
field (RMF) method we calculate an effective free energy $\mathcal{F}_{eff}(S)$ close to the order-disorder transition 
where $S$ is small. We consider density fluctuations  $\delta \wp$ and neglect order parameter fluctuations. The effective 
free energy is
\begin{equation}
\mathcal{F}_{eff}\left(S\right)=-\frac{b_{2}}{2}S^{2}-\frac{b_{3}}{3}S^{3}+\frac{b_{4}}{4}S^{4}
\label{eqfenergy}
\end{equation}
where $b_{2}=\alpha_{1}(\wp)+\alpha_{1}^{\prime}c$, where $c$ is a constant. $\alpha_1^{\prime} = {\partial \alpha_1/\partial \wp|}_{\wp_0}$,
$b_{3}=\frac{a_{0}\wp_0\alpha_{1}^{\prime}}{2D_{\wp}}$, and $b_{4}=\frac{1}{2}\wp_0^2\alpha_{2}$. Both $b_3$ and $b_4$ are positive. 
A detailed calculation for $\mathcal{F}_{eff}$ is shown in Appendix \ref{appRMF}. The density fluctuations introduce a new 
cubic order term in the free energy $\mathcal{F}_{eff} (S)$ that is proportional to the activity strength $a_0$. The presence of 
such term produces a jump $\Delta S = S_{c}=\frac{2b_{3}}{3b_{4}} $ at a density $\wp_c = \wp_{IN}(1-\frac{2b_3^2}{9b_4} ) < \wp_{IN}$. 
Fluctuation in density produces a jump in order parameter and shifts the critical density. Such type of fluctuation induced transitions 
are called fluctuation dominated first order phase transitions in statistical mechanics \cite{coleman} and are widely studied for many
systems \cite{fdfopt,fdfopt2}. The jump in  $S$ and the shift in the transition density are proportional to the activity strength $a_0$, and 
for $a_0=0$ we recover the equilibrium transition.

%%%%%%%%%%%%%%%%%%%%%%%%%%%%%%%%%%%%%%%%%%%%%%%%%%%%%%%%%%%%%%%%%%%%%%%%%%%%%%%%%%%%%%%%%%%%%%%%%%%%%%%%%%%%%%%%%%%%%%%%%%%%%%%%

\section*{Discussion}\label{secsummary}
\noindent In our present  work we have introduced a minimal lattice model for the active nematic and study 
different ordering states in the density-temperature plane. A brief summary of the results is as 
follows. In the low density regime, the system is in the disordered isotropic (I) state with short range orientation 
correlation amongst the particles. In the low temperature regime, large density fluctuation in the active system 
induces a first order transition from the isotropic to the banded state with a jump in the scalar order parameter 
at a density lower than the equilibrium isotropic-nematic (I-N) transition density. The linear stability analysis 
of the isotropic state shows an instability for large activity strength in the low temperature regime. Such instability 
governs the band formation at density below the equilibrium I-N transition density. As we further increase density, bands 
vanish and locally ordered patches appear in the inhomogeneous mixed (IM) state. Renormalised mean field calculation 
confirms the jump in the  scalar order parameter and the shift in the transition density. With increasing temperature 
the shift in the transition density and the jump in scalar order parameter decreases, and no bands appear in the system. 
The IM state is a state with coexisting aligned and disordered domains, similar to the coexisting or defect-ordered 
states found in Ref. \cite{ngo, aparnaredner, shimanatcomm, yeomans, juliaarxiv, ozaarxiv, decamp}. 
 
In the high density regime, the active nematic shows switching between the IM (low $S$) and the homogeneous ordered 
(HO, high $S$) states, i.e., the system shows bistability. In the complete filling limit and with excluded volume 
assumption the active model reduces to the equilibrium model. Therefore, the active model 
tends to show a homogeneous nematic state in the high density regime. However, large activity strength makes
the HO state unstable and leads the system to the IM state. This instability in the HO state is similar to the 
earlier studies in Ref. \cite{aparnamarchetti, shimanatcomm}. Ngo et al. \cite{ngo} considered a two dimensional off-lattice 
model for the active nematic without the exclusion constraint. In the low and moderate density regime, they show a homogeneous 
disordered phase and an inhomogeneous chaotic phase, which are similar to the isotropic and the IM states, respectively.
Similar to their study, the spanning area of the IM state (golden regime in the phase diagram Fig. \ref{fig:phase_snap}(a)) 
along the density axis decreases with the increasing temperature. In the high density limit, they note a homogeneous 
quasi-ordered phase only, which similar to the HO state in our study. However, we show the bistability between the HO and 
the IM state in this density limit.

In conclusion, our lattice model for the active nematic is a simple one to design and execute numerically, and easy to compare 
with the corresponding equilibrium model. It shows new features like the BS in the low temperature regime and the bistability 
in the high density regime, as well as some of the early characterised states, e.g., the IM state. It also shows many basic 
features of the active nematic like large number fluctuation, long-time decay of orientation correlation, transition from SRO 
isotropic to QLRO nematic state. The shift in the transition density due to activity strength compared to the equilibrium model 
can be tested in experimental systems where activity can be tuned. We expect the emergence of the bistability in the high density 
regime in a two dimensional experimental system composed of apolar particles with finite dimension and high activity strength.
It would be interesting to study the model without volume exclusion. In this study, particle orientation has  continuous 
symmetry of $O(2)$. Therefore, the equilibrium limit of our model is an apolar analogue of the two-dimensional XY-model.
One can also study the model with discrete orientation symmetry as in Ref. \cite{solonprl, solonpre, prlperuani2012} and compare the results 
with the corresponding equilibrium model.

%%%%%%%%%%%%%%%%%%%%%%%%%%%%%%%%%%%%%%%%%%%%%%%%%%%%%%%%%%%%%%%%%%%%%%%%%%%%%%%%%%%%%%%%%%%%%%%%%%%%%%%%%%%%%%%%%%%%%%%%%%%%%%%%

%
%\section*{Additional information}\label{addinform}
%\noindent {\bf Supplementary information} accompanies this paper.
%
%\noindent {\bf Competing financial interests:} The authors declare no competing financial interests.
%
%%%%%%%%%%%%%%%%%%%%%%%%%%%%%%%%%%%%%%%%%%%%%%%%%%%%%%%%%%%%%%%%%%%%%%%%%%%%%%%%%%%%%%%%%%%%%%%%%%%%%%%%%%%%%%%%%%%%%%%%%%%%%%%%%
%\newpage
%
%\begin{titlepage}\centering
%{\bf \Large Supplementary Information}
%\end{titlepage}
%\nopagebreak[4]
%
%\section*{Supplementary figure}
%%%%%%%%%%%%%%%%%%%%%%%%%%%%%%%%%%%%%%%%%%%%%%%%%%%%%%%%%%%%%%%%%%%%%%%%%%%%%%%%%%%%%%%%%%%%%%%%%%%%%%%%%%%%%%%%%%%%%%%%%%%%%%%%

\appendix
%%%%%%%%%%%%%%%%%%%%%%%%%%%%%%%%%%%%%%%%%%%%%%%%%%%%%%%%%%%%%%%%%%%%%%%%%%%%%%%%%%%%%%%%%%%%%%%%%%%%%%%%%%%%%%%%%%%%%%%%%%%%%%%%%%%%%%%%%%%%%%%%%%

\renewcommand\thefigure{\thesection\arabic{figure}}
\setcounter{figure}{0}
\section{Order-disorder transition in the EM}\label{appEM}
\begin{figure}[h]
  \includegraphics[width=0.4\linewidth]{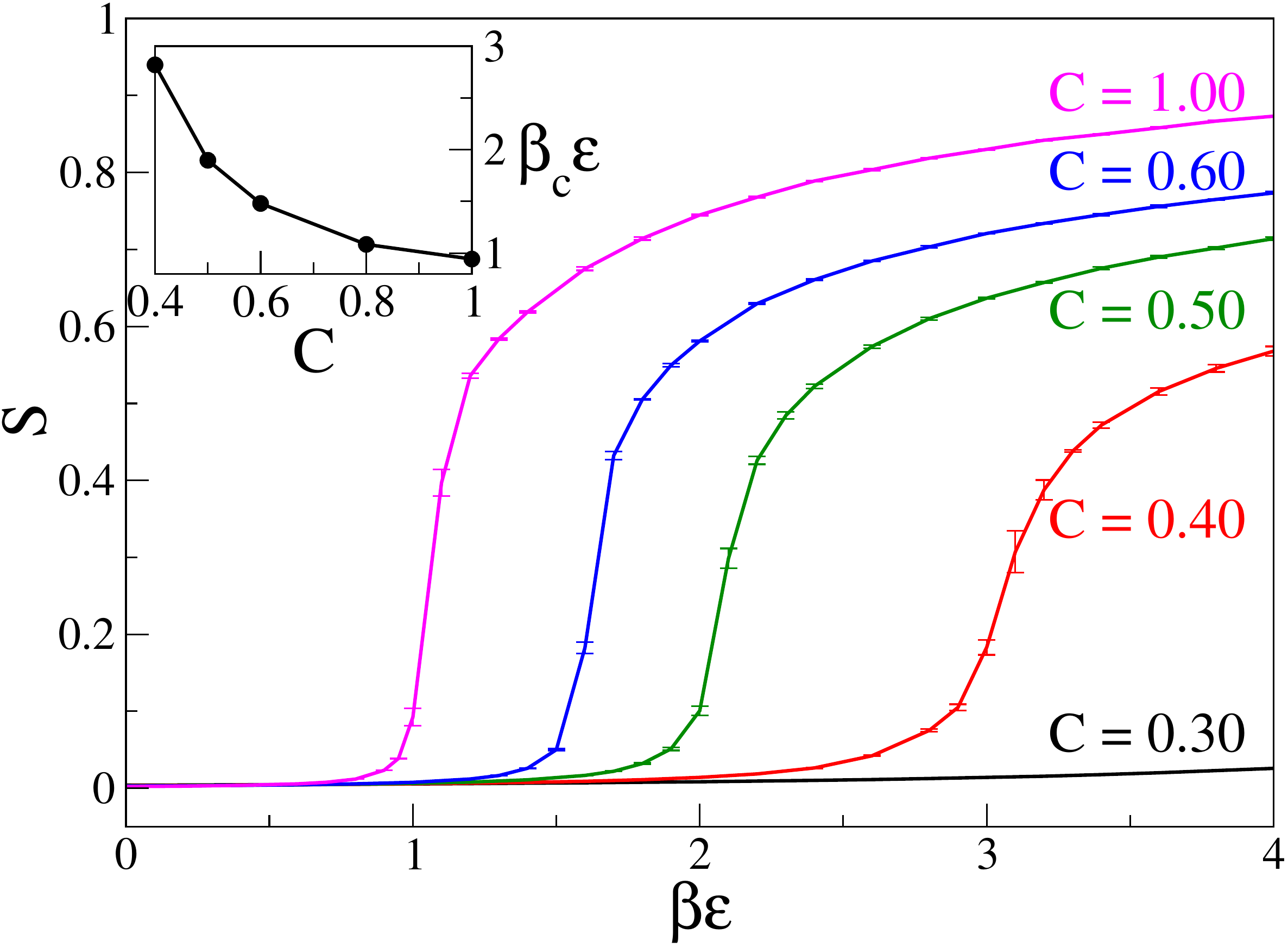}
  \caption{ Order-disorder transition in the equilibrium model. Main - scalar order parameter $S$  versus inverse temperature 
  $\beta\epsilon$ plot for different density $C$. With increasing $\beta\epsilon$ the system goes from the isotropic (small $S$) 
  to the nematic (large $S$) state. 
  Inset - the critical inverse temperature decreases with increasing density. }
\label{figtemp}
\end{figure}

%%%%%%%%%%%%%%%%%%%%%%%%%%%%%%%%%%%%%%%%%%%%%%%%%%%%%%%%%%%%%%%%%%%%%%%%%%%%%%%%%%%%%%%%%%%%%%%%%%%%%%%%%%%%%%%%%%%%%%%%%%%%%%%%%%%%%%%%%%%%%%%%%%

\section{Renormalised mean field (RMF) study of active nematic for small scalar order parameter}\label{appRMF}
\noindent In this section we write an effective renormalised mean field free energy for the scalar order parameter $S$ 
under the small $S$ approximation. We consider the fluctuations in the density and ignore the order parameter fluctuations in 
the coupled hydrodynamic equations of motion for the active nematic. Density fluctuation introduces a cubic 
order term in $S$ in the effective free energy. Such term produces a jump in $S$ at a new transition density $\wp_c$ lower than 
the equilibrium I-N transition point $\wp_{IN}$. Shift in the transition density and the jump $\Delta S$ are directly proportional 
to the activity strength $a_0$. We recover the equilibrium limit for zero $a_0$.

In the main text we write the coupled hydrodynamic equations of motion for the density $\wp$ and the order parameter ${\Pi} = \wp \Q$ 
where nematic order parameter \cite{pgdgenne} is defined as
\begin{eqnarray}
\Q({\bf r}, t)=\frac{S}{2}\left(\begin{array} {c c}
\cos 2\theta ({\bf r}, t) & \sin 2\theta ({\bf r}, t) \\
\sin 2\theta ({\bf r}, t) & -\cos 2\theta ({\bf r}, t)
\end{array} \right)
\label{Q_tensor}
\end{eqnarray}
Here $\theta$ is the coarse-grained orientation at position ${\bf r}$ and time $t$. These hydrodynamic equations are previously 
derived in Ref. \cite{shradhanjop}, but with specific coefficients. Here we retain general coefficients. The density equation is 
a continuity equation $\partial \wp/\partial t = - \nabla  \cdot {\bf J}$, where the current ${\bf J}$ has two parts - active 
and diffusive. Details of these two currents are given in the main text. The activity strength $a_0$ represents the self-propelled 
nature of the particles, $\beta$ is the coupling coefficient of the density in the order parameter equation, $D_{\wp}$ and $D_{\Pi}$  
are the diffusion coefficients in the density and the order parameter equations, respectively. $\alpha_1(\wp)$ and $\alpha_2$ 
represent alignment in the system, and depend on the model parameters. For metric distance interacting model \cite{shradhanjop}, 
$\alpha_1(\wp)$ is a function of density and changes sign at the critical density. We choose $\alpha_1(\wp)= \frac{\wp}{\wp_{IN}}-1$ and 
$\alpha_2 = 1$. Let us consider a small perturbation $\delta\wp$ over the homogeneous steady state solution of the density equation
so that $\wp= \wp_0 + \delta \wp$. Now from the density equation, we obtain
\begin{align}
& a_{0}\nabla_{i}\nabla_{j}{\Pi}_{ij}+D_{\wp}\nabla^{2}\delta \wp = 0 \notag \\
 \Rightarrow \;
& a_{0}\nabla_{j}{\Pi}_{ij}+D_{\wp}\nabla_{i}\delta \wp =\mathbf{c} \equiv constant
\label{eqa3}
\end{align}
where  ${\Pi}_{11} = -{\Pi}_{22} =  \frac{S}{2} \cos(2 \theta)$ and ${\Pi}_{12} = {\Pi}_{21} =  \frac{S}{2} \sin(2 \theta)$. 
Considering only the lowest order terms in $S$ and $\theta$, we obtain
\begin{equation}
\partial_x \delta \wp = -\frac{a_0 \wp_0}{2 D_{\wp}} \partial_x S \Rightarrow \delta \wp(x)=-\frac{a_0 \wp_0}{2 D_{\wp}} S + c_1
\label{eqa4}
\end{equation}
and 
\begin{equation}
 \partial_y \delta \wp = \frac{a_0 \wp_0}{2 D_{\wp}} \partial_y S \Rightarrow \delta \wp(y)=\frac{a_0 \wp_0}{2 D_{\wp}} S + c_2
\label{eqa5}
\end{equation}
Here we assume the system is aligned along one direction, and the variation in orientation is only along the perpendicular direction.
Therefore, we can choose either of equations (\ref{eqa4}) or (\ref{eqa5}). Two constants $c_1$ and $c_2$ are the fluctuations in density when the
nematic order parameter is zero. 

Now from the equation for ${\Pi}_{ij}$, we obtain an effective equation for $S$ as
\begin{equation}
\partial_{t}S=\left\{ \alpha_{1}\left(\wp\right)-\frac{\wp^2}{2}\alpha_{2}S^{2}\right\} S + \mathcal{O}(\nabla^2 S) + \mathcal{O}(\nabla^2 \wp)
\label{eqa6}
\end{equation}
We neglect all the derivative terms and retain only the polynomials in $S$, i.e., we neglect higher order fluctuations.
The Taylor expansion of $\alpha_1(\wp)$ about the mean density $\wp_0$ gives 
$\alpha_1(\wp)=\alpha_1(\wp_{0}+\delta \wp)=\alpha_1(\wp_{0}) + \alpha_1^{\prime} \delta \wp$
where $\alpha_1^{\prime} = \frac{\partial \alpha_1}{\partial \wp}|_{\wp_{0}}$. This gives
\begin{equation}
\partial_{t}S=\left\{ \alpha_{1}\left(\wp_{0}\right)+\alpha_1^{\prime} \delta \wp-\frac{\wp_0^2}{2}\alpha_{2}S^{2}\right\} S
\label{eqa7}
\end{equation}
We can write an effective free energy $\mathcal{F}_{eff}\left(S\right)$ so that
\begin{equation}
\partial_t {S} = -\frac{\delta \mathcal{F}_{eff}(S)}{\delta S}
\label{eqa8}
\end{equation}
Substituting the expression for $\delta \wp$ from equation (\ref{eqa5}), we obtain
\begin{equation}
-\frac{\delta \mathcal{F}_{eff}}{\delta S}  =S\left\{\alpha_{1}\left(\wp_{0}\right)+\alpha_1^{\prime}\left(\frac{a_{0} \wp_0}{2D_{\wp}}S+c_{2}\right)-\frac{\wp_0^2}{2}\alpha_{2}S^{2}\right\}  
\end{equation}
Therefore, 
\begin{equation}
\mathcal{F}_{eff}\left(S\right)=-\frac{b_{2}}{2}S^{2}-\frac{b_{3}}{3}S^{3}+\frac{b_{4}}{4}S^{4}
\label{eqa9}
\end{equation}
where $b_{2}=\alpha_{1}\left(\wp_{0}\right)+\alpha_1^{\prime} c_2$,
$b_{3}=\frac{a_{0}\wp_0\alpha_1^{\prime} }{2D_{\wp}}$
and $b_{4}=\frac{1}{2}\wp_0^2\alpha_{2}$. Since the free energy is a state function, we have assumed the integration 
constant to be zero. Therefore, the fluctuation in the density introduces a cubic order term in the effective free 
energy $\mathcal{F}_{eff}(S)$. Effective free energy in equation (\ref{eqa9}) is similar to the Landau free energy with a 
new cubic order term \cite{chaiklub}. Now we calculate the jump $\Delta S$ and the new critical density from the 
coexistence condition for free energy. Steady state solutions of order parameter 
($S=0$ for isotropic and $S \neq 0$ for ordered state) are given by 
\begin{equation}
\frac{\delta \mathcal{F}_{eff}}{\delta S}=\left(-b_{2}-b_{3}S+b_{4}S^{2}\right)S=0
\end{equation}
Non-zero $S$ is given by $-b_{2}-b_{3}S_c+b_{4}S_c^{2}=0$. Coexistence condition implies
\begin{equation}
\mathcal{F}_{eff}(S_c)=\left(-\frac{b_{2}}{2}-\frac{b_{3}}{3}S_c+\frac{b_{4}}{4}S_c^{2}\right)S_c^{2}=\mathcal{F}_{eff}(S=0)=0
\end{equation}
Hence we get the solution
\begin{equation}
S_{c}=-\frac{3b_{2}}{b_{3}}
\end{equation}
and
\begin{equation}
b_{2}^{c}=-\frac{2b_{3}^{2}}{9b_{4}}
\end{equation}
Therefore, the jump at the new critical point is $\Delta S = \frac{2 b_3}{3 b_4}$.
Since $b_4 >0$ and hence $b_2^c <0$, the  new 
critical density  
\begin{equation}
\wp_{c}=\wp_{IN}\left(1-\frac{2b_{3}^{2}}{9b_{4}}\right) < \wp_{IN}
\label{eqrhoc}
\end{equation}
is shifted to a lower density in comparison to the equilibrium transition density $\wp_{IN}$.
Equation (\ref{eqrhoc}) gives	 the expression for new transition density as given in the main text.
Therefore, using renormalised mean field theory we find a  jump $\Delta S$ at a lower density
as compared to the equilibrium I-N transition density. 

%%%%%%%%%%%%%%%%%%%%%%%%%%%%%%%%%%%%%%%%%%%%%%%%%%%%%%%%%%%%%%%%%%%%%%%%%%%%%%%%%%%%%%%%%%%%%%%%%%%%%%%%%%%%%%%%%%%%%%%%%%%%%%%%%

\begin{acknowledgements}
\noindent We thank Zoltan G. Soos for helpful discussions. S. M. acknowledges Thomas Niedermayer for useful discussions. 
S. M. and M. K. acknowledge financial support from the Department of Science and Technology, India, under INSPIRE award 2012 
and Ramanujan Fellowship, respectively.
\end{acknowledgements}

%%%%%%%%%%%%%%%%%%%%%%%%%%%%%%%%%%%%%%%%%%%%%%%%%%%%%%%%%%%%%%%%%%%%%%%%%%%%%%%%%%%%%%%%%%%%%%%%%%%%%%%%%%%%%%%%%%%%%%%%%%%%%%%%

\section*{Author contributions}\label{authorcontri}
\noindent R.D., M.K. and S.M. designed the research, discussed the results and prepared the manuscript.
R.D. performed simulations.

%%%%%%%%%%%%%%%%%%%%%%%%%%%%%%%%%%%%%%%%%%%%%%%%%%%%%%%%%%%%%%%%%%%%%%%%%%%%%%%%%%%%%%%%%%%%%%%%%%%%%%%%%%%%%%%%%%%%%%%%%%%%%%%%%
%%%%%%%%%%%%%%%%%%%%%%%%%%%%%%%%%%%%%%%%%%%%%%%%%%%
% BIBLIOGRAPHY
%%%%%%%%%%%%%%%%%%%%%%%%%%%%%%%%%%%%%%%%%%%%%%%%%%%

% Create the reference section using BibTeX:

%
%\begin{thebibliography}{10}
%%
%\bibitem{pgdgenne} de Gennes, P. G. \& Prost, J. {\it The Physics of Liquid Crystals} (Oxford: Clarendon Press, 1995).
%%
%\bibitem{shradhanjop} Bertin, E. {\it et al.} Mesoscopic theory for fluctuating active nematics. {\it New J. of Phys.} {\bf 15,} 085032 (2013).
%%
%\bibitem{chaiklub} Chaikin, P. M. \& Lubensky, T. C. {\it Principles of Condensed Matter Physics} (Cambridge: Cambridge University Press, 2000).
%%
%\end{thebibliography}
%
%%%%%%%%%%%%%%%%%%%%%%%%%%%%%%%%%%%%%%%%%%%%%%%%%%%%%%%%%%%%%%%%%%%%%%%%%%%%%%%%%%%%%%%%%%%%%%%%%%%%%%%%%%%%%%%%%%%%%%%%%%%%%%%%%

\end{document}